\RequirePackage{fix-cm}
\documentclass[twocolumn]{svjour3}          
\smartqed  
\usepackage{appendix}
\usepackage{amsmath}
\usepackage{graphicx}
\usepackage{lineno}
\usepackage{array}
\usepackage{longtable}
\usepackage[numbers]{natbib}
\usepackage{tabularx}
\usepackage{booktabs}
\usepackage[caption=false]{subfig}
\usepackage[misc,geometry]{ifsym} 
\usepackage{lipsum}
\newcommand*\patchAmsMathEnvironmentForLineno[1]{%
\expandafter\let\csname old#1\expandafter\endcsname\csname #1\endcsname
\expandafter\let\csname oldend#1\expandafter\endcsname\csname end#1\endcsname
\renewenvironment{#1}%
{\linenomath\csname old#1\endcsname}%
{\csname oldend#1\endcsname\endlinenomath}}%
\newcommand*\patchBothAmsMathEnvironmentsForLineno[1]{%
\patchAmsMathEnvironmentForLineno{#1}%
\patchAmsMathEnvironmentForLineno{#1*}}%
\AtBeginDocument{%
\patchBothAmsMathEnvironmentsForLineno{equation}%
\patchBothAmsMathEnvironmentsForLineno{align}%
\patchBothAmsMathEnvironmentsForLineno{flalign}%
\patchBothAmsMathEnvironmentsForLineno{alignat}%
\patchBothAmsMathEnvironmentsForLineno{gather}%
\patchBothAmsMathEnvironmentsForLineno{multline}%
}

\begin{document}

\title{New Bag of Deep Visual Words based Features to Classify Chest X-ray Images for COVID-19 diagnosis
}


\author{Chiranjibi Sitaula \Letter   \and
        Sunil Aryal 
}


\institute{C. Sitaula \and S. Aryal \at
                School of Information Technology, Deakin University\\
              75 Pigdons Rd, Waurn Ponds VIC 3216 \\
              \email{\{c.sitaula, sunil.aryal\}@deakin.edu.au}           
}
\date{Received: DD Month YEAR / Accepted: DD Month YEAR}

\maketitle

\begin{abstract} {\bf Purpose: }Because the infection by Severe Acute Respiratory Syndrome Coronavirus 2 (COVID-19) causes the pneumonia-like effect in the lung, the examination of Chest X-Rays (CXR) can help diagnose the disease. For automatic analysis of images, they are represented in machines by a set of semantic features. Deep Learning (DL) models are widely used to extract features from images. General deep features extracted from intermediate layers may not be appropriate to represent CXR images as they have a few semantic regions. Though the Bag of Visual Words (BoVW)-based features are shown to be more appropriate for different types of images, existing BoVW features may not capture enough information to differentiate COVID-19 infection from other pneumonia-related infections. {\bf Methods: }In this paper, we propose a new BoVW method over deep features, called Bag of Deep Visual Words (BoDVW), by removing the feature map normalization step and adding the deep features normalization step on the raw feature maps. This helps to preserve the semantics of each feature map that may have important clues to differentiate COVID-19 from pneumonia. {\bf Results: }We evaluate the effectiveness of our proposed BoDVW features in CXR image classification using Support Vector Machine (SVM) to diagnose COVID-19. Our results on four publicly available COVID-19 CXR image datasets reveal that our features produce stable and prominent classification accuracy, particularly differentiating COVID-19 infection from other pneumonia. {\bf Conclusion: } Our method could be a very useful tool for the quick diagnosis of COVID-19 patients on a large scale.
\keywords{
 Bag of Visual Words (BoVW) \and  Bag of Deep Visual Words (BoDVW) \and Chest X-Ray \and  COVID-19\and Deep Features \and SARS-CoV-2 
 }
\end{abstract}

\section{Introduction}
The disease caused by Severe Acute Respiratory Syndrome Coronavirus 2 (SARS-CoV-2) \cite{lai2020severe,li2020game,sharfstein2020diagnostic}, commonly known as COVID-19, was originated in Wuhan city of China in late 2019 \cite{singhal2020review}. It is believed to be originated from bats \cite{CoVIDOrigin_Latine2020,CoVIDOrigin_Nguyen2020}. The virus has been transmitting from human to human all around the world \cite{holshue2020first, giovanetti2020first,bastola2020first}. It has spread over 200 countries in the world at present and become a pandemic that has killed 2,184,120 people\footnote{https://www.worldometers.info/coronavirus/ (accessed date: 28/01/2021)} and 909 people in Australia alone\footnote{https://www.health.gov.au/news/health-alerts/novel-coronavirus-2019-ncov-health-alert/coronavirus-covid-19-current-situation-and-case-numbers. (accessed date: 28/01/2021)}, so far. While analyzing the effect of the SARS-CoV-2 virus in the human body, it has been known that it causes the pneumonia-like effect in the lungs. Thus, the study of chest x-ray images could be an alternative to a swab test for early quick diagnosis of the COVID-19. An automated chest x-ray (CXR) image analysis tool can be very useful to health practitioners for mass screening of people quickly.

For automatic analysis of images using algorithms, they are represented in machines by a set of semantic features. Large artificial neural networks, also known as Deep Learning (DL) models, are widely used to extract features from images and shown to work well in various types of images \cite{sitaula2020hdf,sitaula2020fusion,sitaula2020content,sitaula2020scene,8085139,narin2020automatic}. A few research studies have used DL models to analyze CXR images for coronavirus diagnosis, too. For instance, two recent works \cite{loey2020within,narin2020automatic} include the fine-tuning approach of transfer-learning on pre-trained DL models such as AlexNet \cite{krizhevsky2012imagenet}, ResNet-18 \cite{he2016deep}, GoogleNet \cite{szegedy2015going}, etc. These methods normally require a massive amount of data to learn the separable features in addition to extensive hyper-parameter tuning tasks. However, most of the biomedical images (e.g., COVID-19 CXR images) are normally limited because of privacy issues. Thus, working on a limited amount of data is always a challenging problem in deep learning (DL) models. Similarly, unlike other types of images, existing feature extraction methods such as GAP (Global Average Pooling) features achieved from pre-trained models may not provide accurate representation for CXR images because of their sparsity (i.e., having fewer semantic regions in them). Also, CXR images of lungs infected by COVID-19 and other pneumonia look similar (i.e., there is a high degree of inter-class similarities). There might be subtle differences at very basic level, which, in our understanding, may be captured using the Bag of Words approach over deep features.

Bag of Visual Words (BoVW)-based features are shown to be more appropriate in images with the characteristics discussed above (sparsity and high inter-class similarity). They consider visual patterns/clues (known as visual words) in each image in the collection, thereby capturing sparse interesting regions in the image, which are useful in dealing with the inter-class similarity problem to some degree. BoVW-based feature extraction approach is popular not only in traditional computer vision-based methods such as Scale Invariant Features Transform (SIFT) \cite{lowe2004distinctive} but also in DL-based methods due to its ability to capture semantic information extracted from the feature map of pre-trained DL models. The Bag of Deep Visual Words (BoDVW) features designed for one domain may not work well for another domain due to the varying nature of the images. For example, the Bag of Deep Convolutional Features (DCF-BoVW) \cite{wan2018dcf} designed for satellite images may not work exactly for biomedical images such as CXR images. This is because of the fact that satellite image contains numerous semantic regions scattered in the image (dense) and thus, DCF-BoVW could capture enough semantic regions of such images. However, the CXR images contain fewer semantic regions (sparse), which may not be captured accurately by DCF-BoVW.

\begin{figure}
    \centering
     \subfloat[DCF-BoVW]{\includegraphics[width=70mm, height=70mm,keepaspectratio]{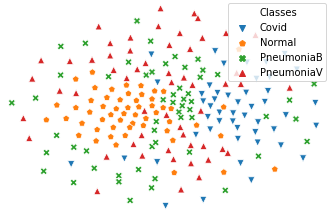}}
 \hspace{0pt}
 \subfloat[Our method]{\includegraphics[width=70mm, height=70mm,keepaspectratio]{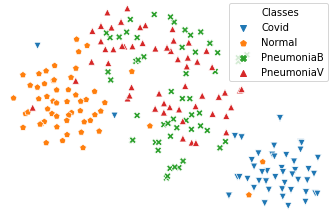}}
    \caption{Scatter plot of two dimensional projection of features produced by DCF-BoVW and our proposed method based on t-SNE visualization on chest x-ray images of Dataset 4 \cite{cohen2020COVID,kermany2018identifying}. 
    }
    \label{fig:my_label}
\end{figure}

In this paper, we propose a new BoDVW-based feature extraction method to represent CXR images. Our method eliminates some of the intermediate steps present in DCF-BoVW \cite{wan2018dcf} and adds new steps because of the nature of CXR images. For this, we adopt the following steps. First, we extract the raw feature map from the mid-level ($4^{th}$ pooling layer) of the VGG16 pre-trained DL model \cite{simonyan2014very} for each input image. We prefer the $4^{th}$ pooling layer in our work, which has been chosen by empirical study and suggestion from recent work by Sitaula et al. \cite{sitaula2020attention}.
Next, we perform L2-normalization of each deep feature vector over the depth of the feature map. Using the training set, we design a codebook/dictionary over such deep features extracted from all the training images. Next, based on the codebook, we achieve our proposed features using a bag of visual words method for each input image. Last, such features based on the bag of visual words method is normalized by L2-norm, which acts as the final representation of the input image. Because our final features are based on patterns extracted from mid-level features from training images, they capture the more discriminating clues of sparse CXR images. The comparison of two-dimensional projections of features produced by DCF-BoVW and our proposed method on the COVID-19 image dataset \cite{cohen2020COVID} based on the t-SNE visualization \cite{maaten2008visualizing} is shown in Fig. \ref{fig:my_label}. It reveals that our features impart the higher separability among different classes.

The main {\bf contributions} in our work are listed below:
\begin{enumerate}
    \item[(a)] Propose to use the improved version of a bag of visual words method over deep features to work for the covid-19 CXR image representation. 
    \item[(b)] Analyze the classification performance of our method across deep features extracted from five different pooling layers of the VGG16 model. Due to higher discriminability of deep features extracted from mid-level VGG16 model (see details in Sec. \ref{pool_cluster} and Sitaula et al. \cite{sitaula2020attention}), we leverage fourth pooling layer ($p\_4$) for feature extraction in our work. To design a codebook from deep features in our work, we use unsupervised clustering with the simple $k$-means algorithm.
    \item[(c)] Evaluate our method on four datasets against the state-of-the-art methods based on pre-trained DL models in the covid-19 CXR classification task using the Support Vector Machine (SVM) classifier. 
    The results show that our method produces stable and state-of-the-art classification performance.
\end{enumerate}

The remainder of the paper is organized as follows. In Sec. \ref{related_works}, we review some of the recent related works on CXR image representation and classification. Similarly, we discuss our proposed method in Sec. \ref{proposed_method} in a step-wise manner. Furthermore, Sec. \ref{experiment} details the experimental setup, performance comparison, and ablative study associated with it. Finally, Sec. \ref{conclusion} concludes our paper with potential directions for future research.

\section{Related works}
\label{related_works}
 Deep Learning (DL) has been a breakthrough in image processing producing significant performance improvement in tasks such as classification, object detection, etc. A DL model is a large Artificial Neural Network (ANN), which has been designed based on the working paradigm of brain. If we design our DL model from scratch and train it, it is called a user-defined DL model. Similarly, if we use existing deep learning architectures pre-trained on large datasets, such as ImageNet \cite{deng_imagenet:_2009} or Places \cite{zhou2016places}, they are called pre-trained DL models. The features extracted from intermediate layers of DL models, either user-defined or pre-trained, provide rich semantic features to represent images that result in significantly better task-specific performance than traditional computer vision methods such as Scale Invariant Feature Transform (SIFT) \cite{lowe2004distinctive}, Generalized Search Tree (GIST)-color \cite{oliva_modeling_2001}, Generalized Search Trees (GIST) \cite{oliva2005gist}, Histogram of Gradient (HOG) \cite{dalal2005histograms}, Spatial Pyramid Matching (SPM) \cite{lazebnik2006beyond}, etc. 
 
 Thus, in this section, we review some of the recent works in chest x-ray classification using DL models \cite{stephen2019efficient,islam2019automatic,ayan2019diagnosis,varshni2019pneumonia,chouhan2020novel,loey2020within,sasaki2012ensemble,narin2020automatic,ozturk2020automated,luz2020towards,panwar2020application,sitaula2020attention}. 
We categorize them into two groups: \ref{sub_sec1} standalone deep learning algorithms and \ref{sub_sec2} ensemble learning algorithms

\begin{figure*}[htb]
    \centering
    \includegraphics[width=\textwidth, height=130mm,keepaspectratio]{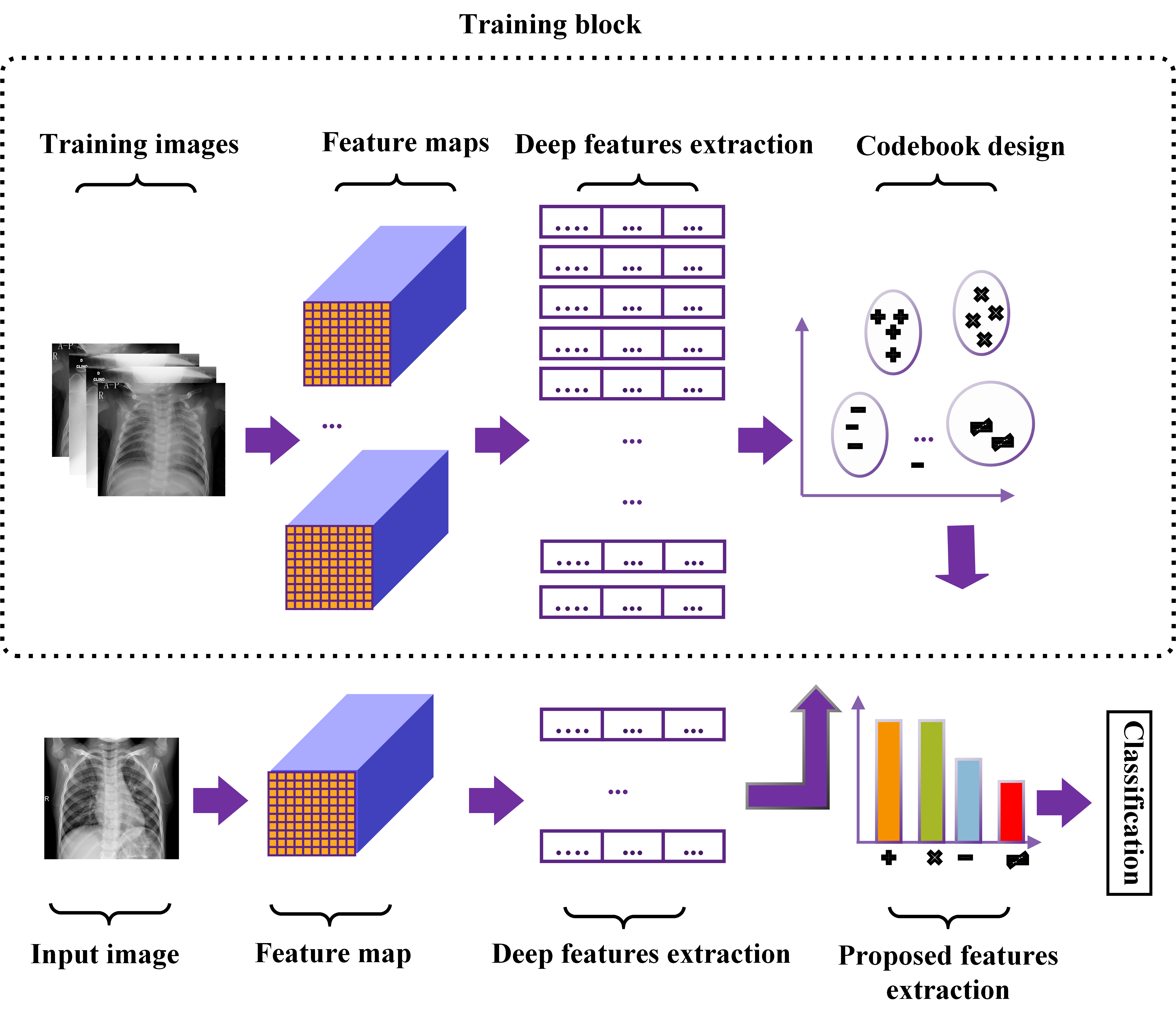}
    \caption{The overall pipeline of the proposed method. Based on the codebook/dictionary achieved from training block, the proposed features vector is extracted for each input image using the bag of visual features approach.
    }
    \label{fig:pipeline}
\end{figure*}

\subsection{Standalone deep learning algorithms}
\label{sub_sec1}
At first, Stephen et al. \cite{stephen2019efficient} presented a new model for the detection of pneumonia using DL and machine learning approach. They trained a Convolutional Neural Network (CNN) from scratch using a collection of CXR images.
Islam et al \cite{islam2019automatic} devised a Compressed Sensing (CS)-based DL model for the automatic classification of CXR images for pneumonia disease.
Similarly, Ayan et al. \cite{ayan2019diagnosis} used DL models on CXR images for early diagnosis of pneumonia. They used Xception \cite{chollet2017xception} and VGG16 \cite{simonyan2014very} pre-trained models. Their results unveil that the VGG16 model outperforms the Xception model in terms of classification accuracy. This strengthens the efficacy of VGG16 model for CXR image representation and classification.
Thus, the use of a pre-trained model became widespread in the representation and classification CXR images. 
For example, Varshni et al. \cite{varshni2019pneumonia} leveraged several pre-trained models such as VGG16 \cite{simonyan2014very}, Xception  \cite{chollet2017xception}, ResNet50 \cite{he2016deep}, DenseNet121 \cite{huang2017densely}, and DenseNet169 \cite{huang2017densely} individually as the features extractors and trained four classifiers separately using SVM \cite{Hearst:1998:SVM:630302.630387}, Random Forest \cite{breiman2001random}, k-nearest neighbors \cite{altman1992introduction}, and Naïve Bayes \cite{lewis1998naive} for the classification purpose.
Furthermore, Loey et al. \cite{loey2020within} used Generative Adversarial Networks (GAN) \cite{goodfellow2014generative} and fine-tuning on AlexNet \cite{krizhevsky2012imagenet}, ResNet18 \cite{he2016deep}, and GoogleNet \cite{szegedy2015going} for the classification of the COVID-19 CXR dataset, where images belong to 4 categories.
In their method, GAN was used to augment the x-ray images to overcome the over-fitting problem during the training phase. 
Moreover, Khan et al. \cite{khan2020coronet} devised a new deep learning model using the Xception \cite{chollet2017xception} model, where they performed fine-tuning using CXR images.

Moreover, Ozturk et al. \cite{ozturk2020automated} established a new deep learning model for the categorization of COVID-19 related CXR images that uses DarkNet19 \cite{redmon2017yolo9000}.
Furthermore, Luz et al. \cite{luz2020towards} devised another novel deep learning (DL) model, which uses the EfficientNet \cite{tan2019efficientnet} model, which adopts transfer learning over CXR images for the classification task.
Furthermore, Panwar et al. \cite{panwar2020application} established a new model, which is called nCOVnet, using the VGG16 model, which imparts a prominent accuracy for COVID-19 CXR image analysis. This further claims that the VGG16 model, which was quite popular in the past, is still popular in CXR image analysis.
Recently, Sitaula et al. \cite{sitaula2020attention} established an attention module on top of the VGG16 model (AVGG) for the CXR images classification. Their method outperforms several state-of-the-art methods.

\subsection{Ensemble learning algorithms}
\label{sub_sec2}
Ensemble learning methods have also been used in CXR image representation and classification where different types of features are combined for better discrimination of images.  
Zhou et al. \cite{zhou2002lung} proposed an ensemble learning approach of several ANNs for the lung cancer cell identification task.
Sasaki et al. \cite{sasaki2012ensemble} established an ensemble learning approach using DL on CXR images. In their method, they performed several filtering and pre-processing operations on images and then ensembled them using DL for the detection of abnormality in CXR images.
Li et al. \cite{li2018false} also utilized multiple CNNs to reduce the false positive results on lung nodules of CXR images.
Moreover, Islam et al. \cite{islam2019automatic} designed an ensemble method to aggregate different pre-trained deep learning models for abnormality detection in lung images.
Chouhan et al. \cite{chouhan2020novel} introduced a model, where the outputs of 5 pre-trained deep learning models, namely AlexNet, ResNet18, DenseNet121, GoogleNet, and Inception-V3, were ensembled for the detection of pneumonia using transfer learning. This helps to learn multiple types of information achieved from various pre-trained DL models to bolster the classification performance. 
Nevertheless, ensemble learning algorithms are arduous for which we need to be vigilant in hyper-parameter tuning in addition to the over-fitting problem.

Most existing methods in the literature need a huge amount of data for fine-tuning DL models and most of them extract high-level features, which may not be sufficient for CXR images. They require mid-level features that are neither more generic nor more specific. In the next section, we introduce our proposed approach to extract such mid-level features.

\section{Proposed method}
\label{proposed_method}

The mid-level features of CXR images can be achieved from the feature maps extracted from the intermediate layers of pre-trained models using a Bag of Visual Words (BoVW) method. Since CXR images are sparse (having few semantic regions), an existing bag of visual words method that has been applied to represent other images (e.g., satellite images) may not work accurately in this domain. To this end, we propose an improved version of a bag of visual words method on deep features to represent CXR images more accurately. In this section, we discuss the steps involved in our proposed feature extraction method. There are three main steps in our method: deep features extraction (Sec. \ref{step_1}), unsupervised codebook (dictionary) design (Sec. \ref{step_2}), and proposed features extraction (Sec. \ref{step_3}). The overall pipeline of the proposed method is shown in Fig. \ref{fig:pipeline}.

\subsection{Deep features extraction}
\label{step_1}
At first, we extract the deep features from the feature map of the $4^{th}$ pooling ($p\_4$) layer from VGG16 \cite{simonyan2014very}, which is a deep learning model pre-trained on ImageNet \cite{deng_imagenet:_2009}. We prefer VGG16 in our work because of three reasons. First, it has a unrivalled performance in recent biomedical image analysis works such as COVID-19 CXR image analysis \cite{sitaula2020attention}, breast cancer image analysis \cite{sitaula2020fusion}, etc. Second, it is easy to analyze and experiment with its five pooling layers. Third, it uses smaller-sized kernels, which could learn distinguishing features of biomedical images at a smaller level.

\begin{figure*}[tb]
\begin{center}
\includegraphics[width=\textwidth, height=90mm,keepaspectratio]{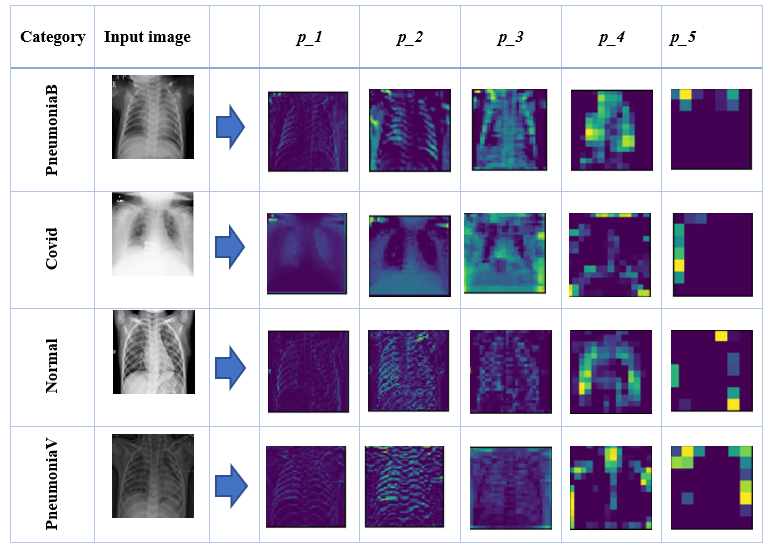}
   \caption{Feature maps of an input image from each of the four categories in the COVID-19 dataset extracted from the five pooling layers of VGG16. $p\_i$ ($i=1,2,\cdots,5$) represents the $i^{th}$ polling layer.}
  \label{fig:feature_maps}
 \end{center}
  \end{figure*}

We believe that $4^{th}$ layer of such a model has a higher level of discriminability than other layers as seen in Fig. \ref{fig:feature_maps}. The detailed discussion about the efficacy of the $4^{th}$ pooling layer is also presented in Sec. \ref{pool_cluster}. 
Furthermore, we use the VGG16 model due to its simple and prominent features extraction capability in various types of image representation tasks \cite{sitaula2020hdf,kumar2020deep,8085139}. 
Authors in \cite{sitaula2020attention,sitaula2020fusion} highlight the importance of $4^{th}$ pooling layer compared to other layers in biomedical imaging for separable feature extraction.
The size of the features map from the $p\_4$ layer of the VGG16 model is $3$-D shape having $H=14$ (height), $W=14$ width, and $L=512$ (length). From each feature map, we achieve $14\times 14$ number of features, each of size $512$. Then, each feature vector is L2-normalized. This normalization helps to preserve the separability of deep features of images \cite{8085139}. 
Let us say that an input image yields feature map with $14 \times 14=196$ number of features vectors that are represented by $x_{0}$, $x_{1}$, $x_{2}$,$\cdots$,$x_{196}$. Each features vector $x_i$ is of $512$-D size (i.e., $|x_i|=512$), which is then normalized by L2-norm as seen in Eq. \eqref{eq:0}.

\begin{equation}
{x'_{i}}=   \frac{x_{i}}{||x_{i}||_2+\epsilon}
\label{eq:0}
\end{equation}

In Eq. \eqref{eq:0}, the features vector $x'_{i}$ represents the $i^{th}$ normalized deep features vector extracted from the corresponding feature map. While achieving such features vector, we add $\epsilon=0.00000008$ with denominator to avoid the divide by zero exception because the feature map obtained for chest x-ray images is sparse and it is more likely to encounter the divide by zero exception in most cases.

\subsection{Unsupervised dictionary (codebook) design}
\label{step_2}
We used deep features (extracted from the VGG16 model as discussed above in Sec.~\ref{step_1}) of all training images to design a dictionary or codebook. Each image provides $\{x'_i\}_{i=1}^{196}$ deep features and let's say there are $m$ training images. Thus, the total number of deep features to design our codebook is $196\times m$. To design the codebook or dictionary, we utilize a simple, yet popular unsupervised clustering algorithm called $k$-means \cite{Jin2010} that groups deep features having similar patterns into clusters. Given a parameter $k$, $k$-means provide $k$ groups or clusters ($\{c_1, c_2, \cdots, c_k\}$) of deep features where deep features in each group are similar (i.e., they capture similar patterns of images). We use such $k$ cluster centroids as a dictionary or codebook of deep visual words which is used to extract features for each input image.

\begin{figure}[b]
\begin{center}
 \subfloat[]{\includegraphics[width=25mm, height=25mm]{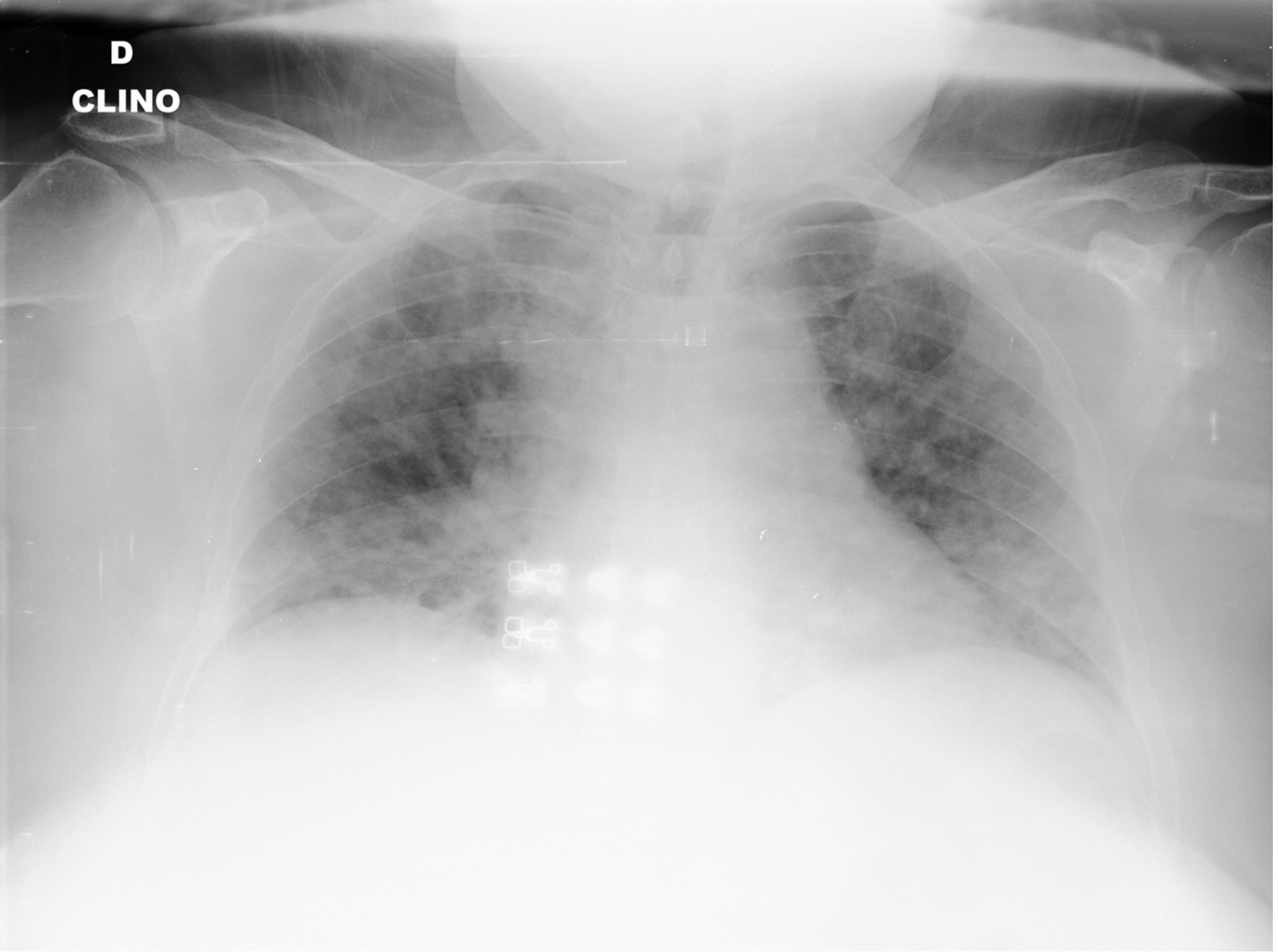}}
 \hspace{5pt}
 \subfloat[]{\includegraphics[width=25mm, height=25mm]{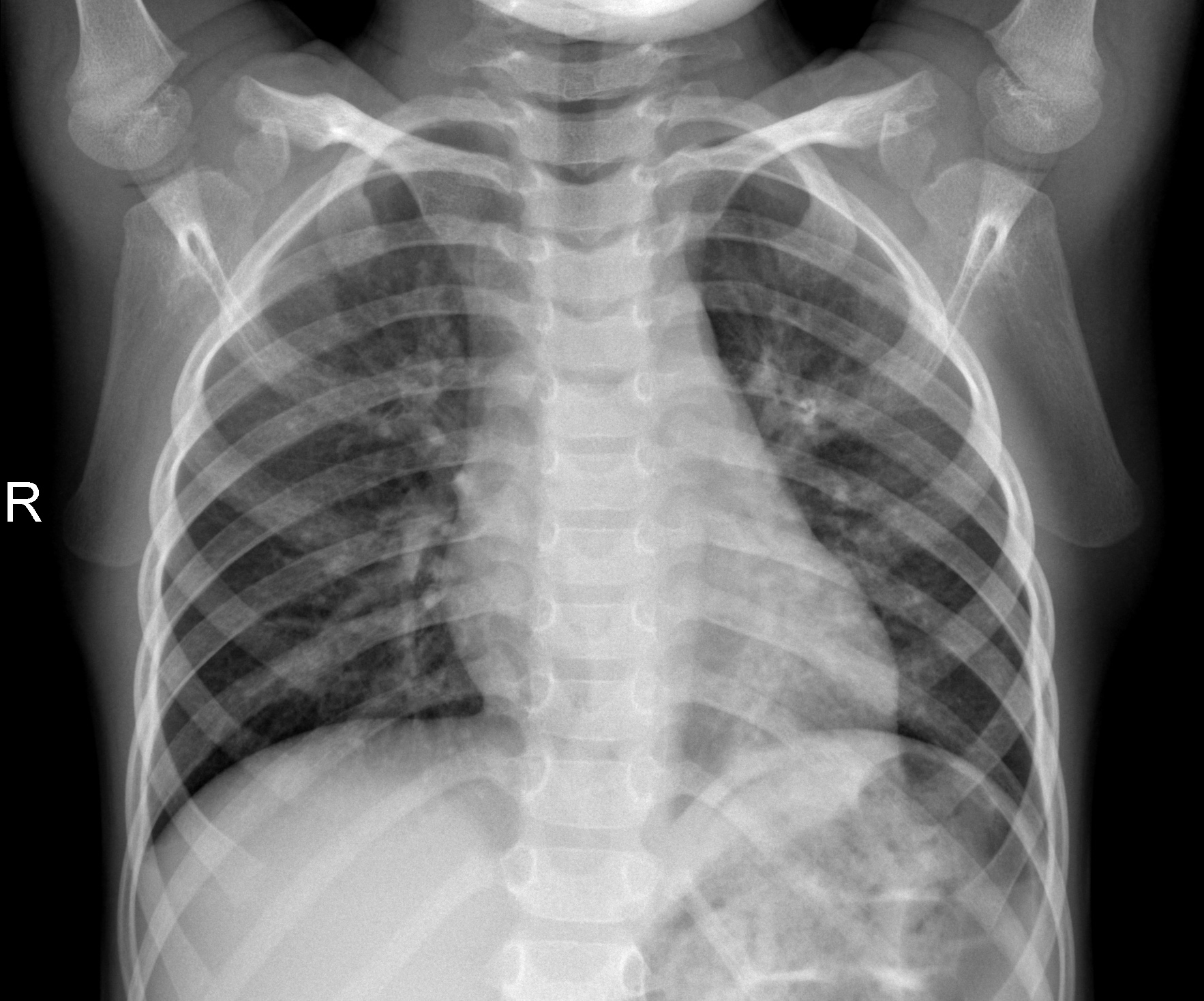}}
  \hspace{5pt}
 \subfloat[]{\includegraphics[width=25mm, height=25mm]{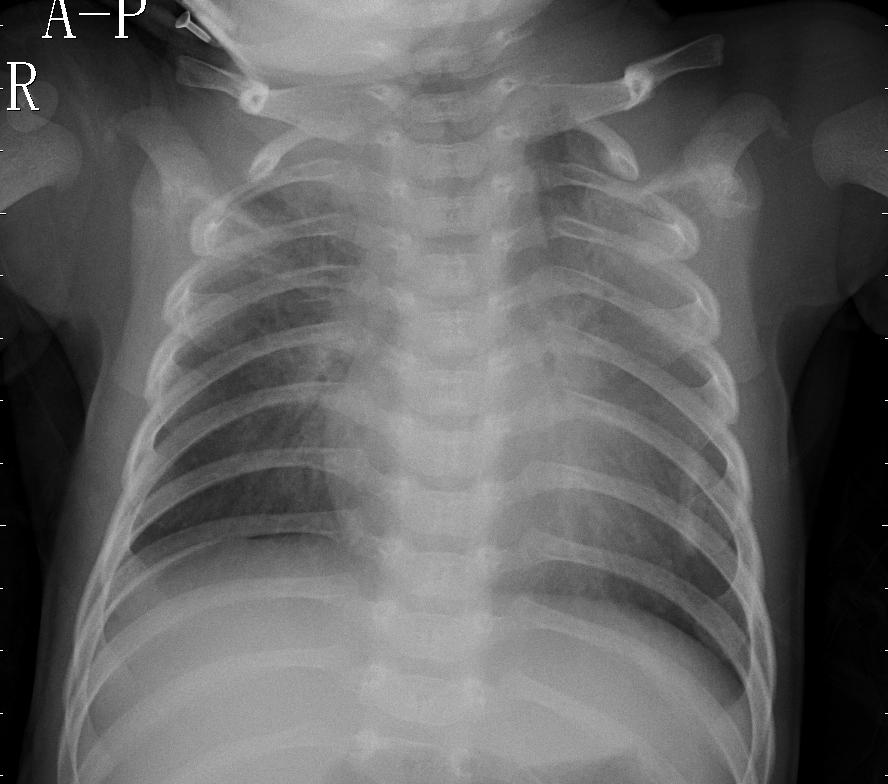}}
  \hspace{5pt}
 \subfloat[]{\includegraphics[width=25mm, height=25mm]{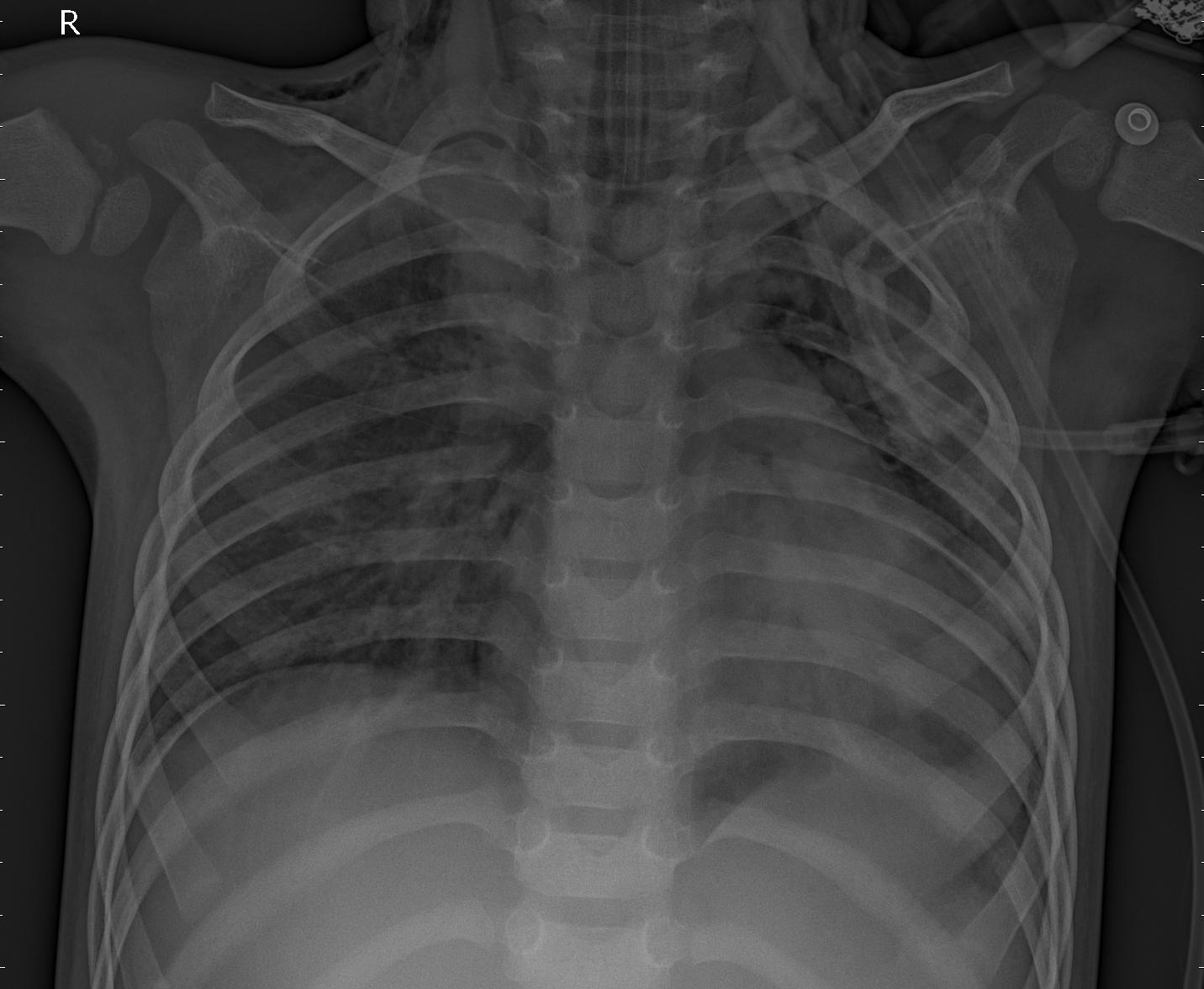}}
   \caption{
   Example images of chest x-ray images from Dataset 4 \cite{cohen2020COVID,kermany2018identifying} for four classes: (a) Covid, (b) Normal, (c) PneumoniaB, and (d) PneumoniaV.
   }
  \label{fig:samples}
 \end{center}
  \end{figure}
  
\subsection{Proposed feature extraction}
\label{step_3}
To extract features of each input image $y$, we first follow step \ref{step_1} to achieve $196$ normalized deep features of $y$ and then, design a histogram based on the dictionary defined in step \ref{step_2}. The size of histogram is $k$ (the dictionary size) where each code (cluster centroid) in the dictionary $c_j$ has a weight $w_j$. All $196$ deep features of $y$ are assigned to their nearest centroids. The weight $w_j$ is the number of deep features assigned to the cluster $c_j$. In other words, histogram is a bag of visual words (centroids) where weights are their frequencies. The resulting features of $y$ is a $k$-D vector $\{w_1,w_2,\cdots,w_k\}$. The extracted bag of visual words features vector is, finally, normalized as in Eq. \eqref{eq:0}, which acts as our proposed features of the corresponding input image.

\subsection{Difference between our BoVW and DCF-BoVW features}
The main differences between our BoVW and DFC-BoVW features are explained in three different aspects.

Firstly, the L1-normalisation used by the DCF-BoVW method is more suitable for dense images such as satellite images. However, since the chest x-ray images are sparse in nature, such normalization becomes counterproductive as it masks some discriminating clues. Thus, we eliminate this normalization in our method due to the nature of chest x-ray images.

Secondly, we apply L2-normalisation to the deep features extracted from the unnormalized feature maps to exploit the property of cosine similarity in the $k$-means clustering. Note that Euclidean distance on the L2-normalised feature is equivalent to using cosine distance. The directions of deep features are more important than their lengths to group vectors with similar patterns into clusters to define our codebook. This will help us to detect sparse patterns in images which can be useful in discriminating abnormalities in x-ray images. 

Finally, we replace the L1-normalisation of the final BoVW features used in the DCF-BoVW method by L2-normalisation. Again, this allows us to exploit the property of cosine similarity in the SVM's RBF kernel. Because BovW features are sparse as many vector entries are zeros, cosine similarity is more appropriate than the Euclidean distance.

\section{Experimental setup and comparison}
\label{experiment}
\subsection{Dataset}

We utilize 4 COVID-19 CXR image datasets 
that are publicly available.  To evaluate our method on such datasets, we divide the images of each dataset into a 70:30 ratio for the train:test set for each category. We utilize the average accuracy of five different runs to present in the table for the comparison purpose. 

{\bf Dataset 1 }\cite{ozturk2020automated} comprises 3 categories: Covid-19, Pneumonia, and No\_findings. Here, each category has at least 125 images. The No\_findings category comprises several ambiguous and challenging CXR images.

{\bf Dataset 2 }\cite{khan2020coronet} comprises 4 categories: Covid, Normal, Pneumonia Viral (PneumoniaV) and Pneumonia Bacteria (PneumoniaB)

{\bf Dataset 3 }\cite{khan2020coronet,ozturk2020automated} includes 5 categories: Covid, No\_findings, Normal,  Pneumonia Bacteria (PneumoniaB), and Pneumonia Viral (PneumoniaV). Dataset 3 is the combination of No\_finding category from Dataset 1 and other categories from Dataset 2. Here, each category includes at least 320 CXR images.

{\bf Dataset 4}\cite{cohen2020COVID,kermany2018identifying} has 4 categories: Covid, Normal, PneumoniaV, and PneumoniaB, where each category contains at least $69$ images. This dataset has been
used by \cite{loey2020within}, which can be downloaded from the link \footnote{COVID-19 Dataset Available online: https://drive.google.com/uc?id=1coM7x3378f-Ou2l6Pg2wldaOI7Dntu1a (accessed on Apr 17, 2020).}

Example images of covid-19 are shown in Fig. \ref{fig:samples}. Also, further detailed information of all datasets are provided in Table \ref{tab:dataset_description}.

\begin{table}[tb]
    \centering
\caption{Detailed description of datasets used in our work}
    \begin{tabular}{p{1cm}|p{1cm}|p{3cm}|p{1cm}}
    \toprule
    Dataset   & \# of images & Categories  & Ref.\\
    \midrule
       Dataset 1 (D1)  & 1,125 & Covid-19, Pneumonia, No\_findings & \cite{ozturk2020automated}\\
       Dataset 2 (D2)  &1,638 & Covid, Normal, PneumoniaB, PneumoniaV &\cite{khan2020coronet}\\
        Dataset 3 (D3)  &2,138 & Covid, Normal, No\_findings, PneumoniaB, PneumoniaV &\cite{khan2020coronet,ozturk2020automated}\\
        Dataset 4 (D4) &320 &Covid, Normal, PneumoniaB, PneumoniaV &\cite{cohen2020COVID,kermany2018identifying} 
        \\
       \midrule
    \end{tabular}
    \label{tab:dataset_description}
\end{table}

\subsection{Implementation}
To implement our work, we use Keras \cite{chollet2015keras} implemented in Python \cite{Rossum:1995:PRM:869369}. Keras is used to implement the pre-trained model in our work. We use the number of clusters $k=400$ in $k$-means clustering to define the dictionary to extract proposed features. For the classification purpose, we use a Support Vector Machine (SVM) classifier implemented using Scikit-learn \cite{pedregosa2011scikit} in Python.
We normalize and standardize our features to feed into the SVM classifier. 
Moreover, we fix the kernel as radial basis function ($RBF$) and $\gamma$ parameter as $1e-05$ in SVM. We automatically tune the cost parameter $C$ in the range of $\{1,10,20,\cdots, 100\}$ on the training set using a $5$-fold cross-validation method and use the optimal setting to train the model using the entire training set and test on the test set. We execute all our experiments on a workstation with NVIDIA Geforce GTX 1050 GPU and 4 GB RAM.

\begin{table}[t]
    \centering
\caption{Comparison with previous methods on four datasets (D1, D2, D3, and D4) using average classification accuracy (\%) over five runs. Note that - represents the unavailable accuracy because of the over-fitting problems in existing DL-based methods using transfer learning on D4.}
    \begin{tabular}{p{2cm}|p{1cm}|p{1cm}|p{1cm}|p{1cm}}
    \toprule
    Method & D1 (\%)& D2 (\%) & D3 (\%)&D4 (\%)\\
    \midrule
       DCF-BoVW, 2018 \cite{wan2018dcf} &75.31 &81.53 &83.72 &72.46 \\
       CoroNet, 2020 \cite{khan2020coronet} & 76.82&80.60&83.41 &-\\
       Luz et al., 2020 \cite{luz2020towards} &47.51 &84.29&79.96 &- \\
       nCOVnet, 2020 \cite{panwar2020application} &62.95 &70.62&67.67 &- \\
       AVGG, 2020 \cite{sitaula2020attention} & 79.58& 85.43& 87.49&- \\
     \midrule
    \bf Ours & \bf82.00 & \bf 87.86 & \bf 87.92 &\bf 83.22 \\
       \bottomrule
    \end{tabular}
    \label{tab:existing_methods}
\end{table}

\subsection{Comparison with state-of-the-art methods}
We present the results of the experiments conducted to compare our method with five recent state-of-the-art methods (one method uses the BoW approach over deep features and four methods adopt transfer-learning approach) that are based on pre-trained models on four CXR image datasets (D1, D2, D3, and D4) in Table \ref{tab:existing_methods}. In the table, the second, third, fourth, and fifth columns enlist the accuracies of contending methods in D1, D2, D3, and D4, respectively. Note that the accuracies reported in the table are averaged accuracy of five runs for each method.

Results in the second column of Table \ref{tab:existing_methods} show that our method outperforms all five contenders with the accuracy of 82.00\% on D1. This further highlights that it imparts the performance increment of at least 2.50\% from the second-best method (AVGG \cite{sitaula2020attention}) and at least 40\% accuracy from the worst method (Luz et al. \cite{luz2020towards}). Similarly, on D2 in the third column of Table \ref{tab:existing_methods}, we notice that our method outperforms all five methods with an accuracy of 87.86\%, which is at least 2.43\% higher than the second-best method (AVGG \cite{sitaula2020attention}) and at least 17\% higher than the worst-performing method (nCOVnet \cite{panwar2020application}). In the fourth column of Table \ref{tab:existing_methods} on D3, we observe that our method, which yields 87.92\% accuracy, is superior to the second-best method (AVGG \cite{sitaula2020attention}) with a slim margin of 0.43\%, whereas it imparts over 20\% accuracy against the worst performing method (nCOVnet \cite{panwar2020application}). Last but not the least, in the fifth column of Table \ref{tab:existing_methods} on D4, we notice that our method, which produces 83.22\%, outperforms the DCF-BoVW \cite{wan2018dcf} with the margin of over 10\% accuracy. Please note that for D4, we only compare our method with DCF-BoVW \cite{wan2018dcf}, which can work for a limited amount of data, only and do not compare with other DL-based methods that uses transfer learning because this dataset has a very limited number of CXR images.

The comparison of our method against five different recent DL-based methods on four datasets unveils that our method provides a stable and prominent performance. This result further underscores that the classification performance of the bag of words approach, which capture the more detailed spatial information of deteriorated regions more accurately than other methods, seems more appropriate to CXR image analysis (e.g., COVID-19 CXR images) than other DL-based methods using transfer learning approach. 

\begin{figure}[tb]
\begin{center}
\includegraphics[width=0.45\textwidth, height=90mm,keepaspectratio]{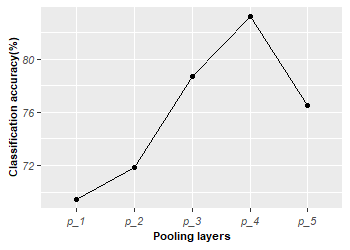}
   \caption{Average classification accuracy (\%) achieved by our method on D4 using deep features extracted from the five pooling layers ($p\_1$ to $p\_5$) of the VGG16 model.
   }
  \label{fig:pooling_layers}
 \end{center}
  \end{figure}

\begin{figure}[tb]
\begin{center}
\includegraphics[width=0.45\textwidth, height=90mm,keepaspectratio]{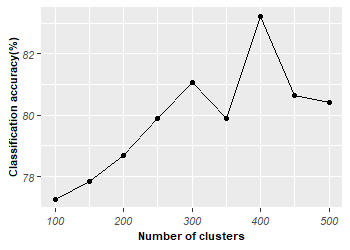}
   \caption{Average classification accuracy (\%) with different cluster number on D4. Note that deep features from the $4^{th}$ pooling layer ($p_4$) were used.}
  \label{fig:cluster_layers}
 \end{center}
  \end{figure}
 
\subsection{Ablative study of pooling layers}
\label{pool_cluster}
In this subsection, we present the results of an ablative study on D4, which is the smallest dataset, to analyze the effect on the classification accuracy of using deep features from the five different pooling layers of VGG16 in our method.
The detailed results are presented in Fig. \ref{fig:pooling_layers}. While observing the line graph, we notice that the $4^{th}$ pooling layer of the VGG16 model produces highly separable features than other pooling layers on the COVID-19 dataset.

\subsection{Ablative study of cluster numbers}
We analyze different number of unsupervised patterns to be used in our experiments on D4. For this, we vary the cluster numbers from $100$ to $500$ using the interval of $50$ and present the results in Fig. \ref{fig:cluster_layers}. From the line graph, we notice that the appropriate number of clusters that produce the best result is $k=400$.

\subsection{Ablative study of class-wise performance}
We study the average class-wise performance of our method on D4. The average class-wise performance are reported using precision, recall, and f1-score, which are defined in Eqs. \eqref{eq:precision},\eqref{eq:recall}, and \eqref{eq:f1-score}, respectively. 
\begin{equation}
    \text{Precision} = \frac{TP}{TP+FP},
\label{eq:precision}
\end{equation}

\begin{equation}
  \text{Recall} = \frac{TP}{TP+FN},
 \label{eq:recall}   
\end{equation}

\begin{equation}
  \text{F1-score} = \frac{2\times(\text{Recall} \times \text{Precision})}{(\text{Recall} + \text{Precision})},
\label{eq:f1-score}
\end{equation}
where $TP$, $FP$, and $FN$ represent true positive, false positive, and false negative results, respectively.
We present the average precision, recall, and f1-score in Table \ref{tab:precision_recall_fscore}. 
The results show the discriminability of our proposed method in all four classes. It shows that our method can distinguish the Covid and normal class well and there is some confusion among two pneumonia classes.

\begin{table}[t] 
\caption{Average class-wise study (\%) over five runs of our method on D4 using precision, recall, and f1-score. 
}
\centering
\begin{tabular}{p{2.5cm} p{1.5cm} p{1.5cm} p{1.5cm}}
\toprule
Class&Precision (\%)& Recall (\%)&F1-score (\%)\\
\midrule
Covid&100.00 &97.20 &98.40 \\
Normal&94.20 &93.60 &93.80 \\
PneumoniaB&75.80 &67.60 &71.00 \\
PneumoniaV&68.00 &76.80 &71.80 \\
\hline
\end{tabular}
\label{tab:precision_recall_fscore}
\end{table}

\section{Conclusion and future works}
\label{conclusion}
In this paper, we propose a new feature extraction method based on Bag of Deep Visual Words (BoDVW) to represent chest x-ray images. Empirical results on the classification of chest x-ray images using the COVID-19 dataset show that our method is more appropriate to represent chest x-ray images. This is mainly because our features can capture a few interesting regions (sparse markers) indicating abnormalities well. Our features are extracted using a visual dictionary defined by the clustering of deep features from all training images. Therefore, they can capture patterns in each training image and thus helps to capture potential markers for various lung infections such as COVID-19 and pneumonia. Also, the size of our proposed features is relatively very small compared to other existing methods and our method runs faster than other existing methods.

Though the evaluation is done on a relatively small dataset, our method shows promising results to detect and distinguish lung infection due to pneumonia and COVID-19. COVID-19 being a relatively new disease and there are not a lot of chest x-ray images available. Nevertheless, given the current crisis with the COVID-19 pandemic, our method which is accurate and fast can be very useful for health professionals for mass screening of people for COVID-19. Accurate detection and distinction of lung infections due to COVID-19 and pneumonia are very important for COVID-19 diagnosis as people infected by these diseases show similar symptoms.  

In the future, it would be interesting to verify our results in a large study with more sample images including other types of lung infection such as tuberculosis. Another potential direction is to investigate if a similar approach can be used to represent other types of medical images such as CT scans, histopathological images, colonoscopy images, etc. 

\bibliographystyle{spbasic}
\bibliography{sample_library}
\end{document}